\documentclass[a4paper]{article}
\usepackage{amsmath}
\usepackage{epsfig}

\begin{document}

\title{2D\ models for MHD flows}
\author{A. Poth\'{e}rat$^a$, J. Sommeria$^b$, \ R. Moreau$^a$\\
$^a$Laboratoire EPM-MADYLAM (CNRS), ENSHMG, B.P. 95,\\ 38402 Saint-Martin-d’H\`eres cedex, France\\
$^b$Laboratoire de physique (CNRS), \'Ecole normale sup\'erieure de Lyon, \\46, all\'ee de l’Italie, 69364 Lyoncedex 07, France
}
\date{21 October 1999}

\maketitle
\textbf{Abstract : }A new model is proposed for low $Rm$ MHD flows which
remain turbulent even in the presence of a magnetic field. These flows
minimize the Joule dissipation because of their tendency to become
two-dimensional and, therefore to suppress all induction effects. However,
some small three-dimensional effects, due to inertia and to the electric
coupling between the core flow and the Hartmann layers, are present even
within the core flow. This new model, which may be seen as an improvement of
the Sommeria-Moreau 2D\ model, introduces this three-dimensionality as a
small perturbation. It yields an equation for the average velocity over the
magnetic field lines, whose solution agrees well with available measurements
performed on isolated vortices.

\textbf{R\'{e}sum\'{e} : }Un nouveau mod\`{e}le est propos\'{e} pour les
\'{e}coulements MHD qui demeurent turbulents, m\^{e}me en pr\'{e}sence d'un
champ magn\'{e}tique. Ces \'{e}coulements minimisent la dissipation par
effet Joule en raison de leur tendance \`{a} devenir bidimensionnels et, par
cons\'{e}quent, \`{a} supprimer tout effet d'induction. Toutefois, la
pr\'{e}sence d'inertie et le couplage \'{e}lectrique entre les couches de
Hartmann et la r\'{e}gion centrale maintient certains effets
tridimensionnels, m\^{e}me en dehors des couches limites. Ce nouveau
mod\`{e}le, qui peut \^{e}tre vu comme un perfectionnement au mod\`{e}le
ant\'{e}rieur de Sommeria-Moreau, introduit cette faible
tridimensionnalit\'{e} comme une petite perturbation. Il conduit \`{a} une
\'{e}quation moyenn\'{e}e sur une ligne de flux magn\'{e}tique, dont la
solution est en bon accord avec des mesures disponibles, effectu\'{e}es sur
des tourbillons isol\'{e}s.

\noindent\makebox[\linewidth]{\rule{\textwidth}{0.4pt}}
\begin{center}
\Large{Mod\`{e}les 2D d'\'{e}coulements MHD}\\
(Version fran\c{c}aise abr\'{e}g\'{e}e)
\end{center}

L'action d'un champ magn\'{e}tique stationnaire et uniforme $\mathbf{B}$ ($%
Oz,$ vertical) sur un \'{e}coulement de fluide conducteur peut se r\'{e}%
sumer \`{a} un ph\'{e}nom\`{e}ne de diffusion qui tend \`{a} effacer les diff%
\'{e}rences de vitesses entre plans horizontaux [1]. En premi\`{e}re
approximation, Lorsque le champ est assez fort, l'\'{e}coulement compris
entre une plaque et une surface libre (distants de $a$) peut \^{e}tre suppos%
\'{e} 2D sauf dans la couche limite voisine de la plaque, o\`{u} les forces
visqueuses sont suffisantes pour \'{e}quilibrer la force de Lorentz et
donner lieu au profil bien connu des couches de Hartmann (d'\'{e}paisseur $%
a/Ha=\left( \rho \nu /\sigma \right) ^{1/2}B^{-1}$). Les \'{e}quations du
mouvement moyenn\'{e}es selon la verticale sont donc particuli\`{e}rement
apropri\'{e}es \`{a} la mod\'{e}lisation de ces \'{e}coulements. Nous nous
proposons d'enrichir ce mod\`{e}le en y int\'{e}grant les effets de
l'inertie.

Dans les \'{e}coulements 2D, chaque terme de la force de Lorentz, d'ordre $%
\sigma B^{2}u$ est grand devant tout autre terme de l'\'{e}quation de Navier
Stokes, si bien que les termes habituels d'inertie ne contr\^{o}lent pas
l'\'{e}coulement 2D (d\'{e}sign\'{e} par $($ $)^{2D}$ dans la suite).Le
courant dans le coeur de l'\'{e}coulement est donc nul dans cette
premi\`{e}re approximation. En fait, un faible courant r\'{e}sulte de
l'action des termes habituels de l'\'{e}quation du\ mouvement (pression,
acc\'{e}l\'{e}ration), petits devant $\sigma B^{2}u$ et qui peuvent donc
\^{e}tre construits sur le profil 2D. En cons\'{e}quence, le courant induit
est 2D et par conservation du courant total, il doit \^{e}tre aliment\'{e}
par un courant vertical lin\'{e}aire en $z,$ qui implique lui-m\^{e}me un
potentiel quadratique en $z$. D'apr\`{e}s l'\'{e}quation (3), le profil de
vitesse corrig\'{e} $\mathbf{u_{\bot }^{\ast }}$ doit lui aussi \^{e}tre
quadratique (6).

Une m\'{e}thode similaire appliqu\'{e}e \`{a} la couche de Hartmann, pour
laquelle le premier ordre r\'{e}sulte de l'\'{e}quilibre entre les forces
visqueuses et les forces de Lorentz conduit \'{e}galement \`{a} un profil
modifi\'{e} (8). Contrairement \`{a} l'approximation classique (sans
inertie), ce nouveau profil induit un d\'{e}bit entre coeur et couche
limite, caract\'{e}ris\'{e} par l'expression (9).

Connaissant le profil vertical complet, il est d\'{e}sormais possible d'int%
\'{e}grer les \'{e}quations du mouvement selon $(Oz)$ et d'en d\'{e}duire
une \'{e}quation d'\'{e}volution pour la vitesse horizontale moyenn\'{e}e $%
\mathbf{\bar{u}}$ entre $z=0$ et $z=a$. Ceci conduit \`{a} l'\'{e}quation
(10), o\`{u} dans un but de simplification, les effets de l'inertie sur le
coeur ne sont pas pris en compte. La r\'{e}solution de cette \'{e}quation
dans la configuration stationnaire et axisymm\'{e}trique d'un tourbillon isol%
\'{e} provoqu\'{e} par une injection de courant ponctuelle situ\'{e}e dans
le plan $z=0$ permet une comparaison avec les exp\'{e}riences de Sommeria
[2] au cours desquelles il a mesur\'{e} des profils radiaux de moment cin%
\'{e}tique pour de telles structures. Les r\'{e}sultats exp\'{e}rimentaux et
analytiques, port\'{e}s sur la figure 1, sont en accord raisonnable : l'\'{e}%
talement des structures constat\'{e} pour des courants d'injection \'{e}lev%
\'{e}s est bien provoqu\'{e} par un d\'{e}bit de fluide \'{e}ject\'{e} de la
couche de Hartmann au centre du vortex (circulation d'Ekman).

\noindent\makebox[\linewidth]{\rule{\textwidth}{0.4pt}}

\section{Introduction.}

It is well known since \cite{Sommeria82} that the action of a steady uniform
magnetic field $\mathbf{B}$ on an electrically conducting fluid
(conductivity $\sigma ,$ density $\rho ,$ kinematic viscosity $\nu $) can be
seen as a diffusion phenomenon which tends to suppress velocity differences
between planes orthogonal to the direction of the magnetic field $\left(
Oz\right) $.\ In the case of a flow bounded by a horizontal plane and a free
surface (separated by a distance $a$), a strong enough vertical magnetic
field makes this phenomenon dominant (even in comparison with inertial
effects) so that the velocity profile does not depend on the $z$ coordinate
except within the Hartmann layer located in the vicinity of the wall
perpendicular to the field where viscous effect are strong enough to balance
the Lorentz force ; the thickness of such layers is $a/Ha$ $=\left( \nu
/\sigma \rho \right) ^{1/2}B^{-1}$. Averaging the motion equations along $\ $%
the $z$ direction yields a 2D model which is well adapted to this remarkable
flow structure and flexible as it allows any model, such as a turbulent one,
for the horizontal flow.

We first review the 2D\ equation \cite{Sommeria82} and then look for 3D
corrections accounting for non linear behavior of both the Hartmann layer
(Ekman-type flows) and the core flow (''barrel'' shaped turbulent eddies).
Implementing the latter into the motion equation averaged along $\left(
Oz\right) $ between $z=0$ and $z=a$ provides an effective 2D model the
predictions of which are in good agreement with experiments performed on
electrically driven isolated vortices \cite{Sommeria88}.

\section{2D models for quasi 2D\ MHD flows.}

In order to describe a weakly 3D flow by a 2D\ equation of motion , the
velocity field is split into its averaged part in the direction of the field 
$\mathbf{\bar{u}}$ and a departure from this value $\mathbf{u}^{\prime }.$
The Navier-Stokes equation integrated along the field direction writes : 
\begin{equation}
(\partial _{t}+\mathbf{\bar{u}.\nabla }_{\bot })\mathbf{\bar{u}}+\overline{%
\mathbf{u}^{\prime }\mathbf{.\nabla }_{\bot }\mathbf{u}^{\prime }}=-\dfrac{1%
}{\rho }\mathbf{\nabla }_{\bot }\bar{p}+\nu \mathbf{\Delta }_{\bot }\mathbf{%
\bar{u}}+\mathbf{\tau }_{W}+\dfrac{\mathbf{u}_{0}}{t_{H}}.
\label{2D integrated eq - general form.}
\end{equation}
It results from the classical Hartmann layer theory (see for instance \cite
{moreau90} p.124-131)\ that an injected current density $j_{W}$ results
in a forcing velocity $\mathbf{u}_{0}=t_{H}B\rho ^{-1}\nabla \Psi _{0}\times 
\mathbf{e}_{z}$ (with $\overline{\mathbf{j}_{\bot }}=\left( {at_{H}}\right)
^{-1}\nabla \Psi _{0}$ and $\Delta _{\bot }\Psi _{0}=-t_{H}\mathbf{j}_{W}$
where $\overline{\mathbf{j}_{\bot }}$ is the vertically averaged horizontal
current density). $t_{H}=a^{2}/\nu 1/Ha$ is the Hartmann damping time which
accounts for Joule and viscous dissipations within the Hartmann layer. The
subscript ( )$_{\bot }$ stands for orthogonal to the field component of
vectors. The Lorentz force has then been expressed in function of mechanical
quantities thanks to a transformation involving the electric current
conservation (\cite{Sommeria82}).

The averaged quantities, the friction terms and the Reynolds-like tensor
appearing in (\ref{2D integrated eq - general form.}) must be derived from a
model of the vertical dependance of the velocity. Sommeria and Moreau (\cite
{Sommeria82}) use the classical profile of the Hartmann layer $\mathbf{u}%
_{h}=\mathbf{u}_{\bot }^{-}\left( 1-\exp (-\xi \right) )$ (where $\xi =z/Ha$%
) near the Hartmann wall and the property that in the approximation of
strong magnetic field ($Ha$ and $N$ both larger than unity , where $%
N=a\sigma B^{2}/(\rho U)$ is the ratio of the Lorentz force and the Inertia
) the electromagnetic action is dominant in the balance of forces in the
core flow. Then, the velocity profile within the core is not dependent upon
the $z$ coordinate (the so called ''2D core model'') and the equation for
the average velocity yields :

\begin{equation}
(\partial _{t}+\mathbf{\bar{u}.\nabla }_{\bot })\mathbf{\bar{u}}=-\dfrac{1}{%
\rho }\mathbf{\nabla }_{\bot }\bar{p}+\nu \mathbf{\Delta }_{\bot }\mathbf{%
\bar{u}}+\dfrac{1}{t_{H}}\left( \mathbf{u}_{0}-\mathbf{\bar{u}}\right) .
\label{SM 82 model}
\end{equation}
The solutions of this equation compare well with explicit linear inertialess
3D solutions in the case of laminar parallel wall side layers (\cite
{Shercliff53}) or isolated vortices aroused by point electrodes \textbf{(}
\cite{Hunt68}). But if inertia becomes important, the comparison with
available measurements suggests that (\ref{SM 82 model}) which is based on a
linear vertical velocity profile, has to account for non-linear effects
adding a 3D\ component to the latter.

\section{3D phenomena.}

\subsection{3D effects in the core flow.}

Strictly 2D\ flows are dominated by the balance between the two components
of the Lorentz force (respectively the induction part, linked with the
motion of the fluid and the electrostatic part) so that no horizontal
electric current occurs in the core and the motion is the result of the
electrostatic force. In order to improve the 2D\ approximation, let us
consider an additional term in this balance $\mathbf{F}_{\bot }$ ( $%
=a/\left( l_{\bot }N\right) \left[ (\partial _{t}+\mathbf{u.\nabla }_{\bot })%
\mathbf{u+}\dfrac{1}{\rho }\mathbf{\nabla }_{\bot }p\right] $ if one wants
to introduce effects of inertia, where $l_{\bot }$ is a typical horizontal
lengthscale ) of smaller order of magnitude which can therefore be assumed
to depend only on the 2D velocity profile : 
\begin{equation}
\mathbf{F}_{\bot }\mathbf{-}\dfrac{1}{\rho }\mathbf{j}_{\bot }\mathbf{\times
B=F}_{\bot }\mathbf{-}\dfrac{\sigma }{\rho }\left( -\mathbf{B}^{2}\mathbf{%
u_{\bot }+\nabla }_{\bot }\phi \times \mathbf{B}\right) =0.
\label{General Balance of forces}
\end{equation}
The right hand side of (\ref{General Balance of forces}) directly comes from
Ohm's law ($\phi $ is the electric potential). This relation shows that the
additional force results in a horizontal current density which is purely
divergent if no flow rate comes from the Hartmann layer (Poth\'{e}rat 
\textit{\ et Al. }\cite{Potherat99}). This current has to be supplied by a
vertical current density coming out from the Hartmann layer which is
obtained using the current conservation : 
\begin{equation}
\sigma \partial _{zz}^{2}\phi =-\mathbf{\nabla }_{\bot }.\mathbf{j}_{\bot
}=\rho \left( \mathbf{\nabla }_{\bot }\times \mathbf{F}_{\bot }\right) .%
\mathbf{B.}  \label{vertcial current density in the core}
\end{equation}
As $\mathbf{j}_{\bot }$ and are $z$ independent, this variation of potential
along the magnetic field lines induces a quadratic dependence upon $z$ for
the vertical profile of velocity in (\ref{General Balance of forces}) : 
\begin{equation}
\partial _{zz}^{2}\mathbf{u}_{\bot }=\dfrac{\rho }{\sigma }\mathbf{\nabla }%
_{\bot }\times \mathbf{\nabla }_{\bot }\times \mathbf{F}_{\bot }.
\label{velocity profile concavity}
\end{equation}
This demonstrates that the velocity cannot be 2D anymore so that the
well-known 2D vortices do not look like columns anymore but rather like
''barrels'' : 
\begin{equation}
\mathbf{u_{\bot }}\left( z\right) \mathbf{=u}_{\bot }^{-}+\dfrac{z}{2}\left(
z-\dfrac{2a}{n}\right) \dfrac{a^{2}}{\nu }\dfrac{N}{Ha^{2}}\mathbf{\nabla }%
_{\bot }\times \mathbf{\nabla }_{\bot }\times \mathbf{F}_{\bot }
\label{barrel effect velocity profile}
\end{equation}
Notice that if no flow rate occurs at the bottom (or the top) of the eddies,
this 3D mechanism induces no additional vertical flow.

\subsection{Inertial effects in the Hartmann layer.}

In the Hartmann layer, the viscous force is of the same order as the Lorentz
force so that the usual exponential leading order profile is already the
result of a balance between both of them and the pressure gradient. In order
to introduce the effects of inertia as a small perturbation, for instance to
account for a big vortex standing over the layer, an additional inertial
term has to be added \ whose $\xi $ dependance comes from this profile. The
motion equation of within such a layer can be written :
\begin{multline}
a/\left( Nl_{\bot }\right) \left[ \partial _{t}+\mathbf{u}_{\bot }^{-}\left(
1-\exp \left( -\xi \right) \right) \mathbf{.\nabla }_{\bot }\right] \mathbf{u%
}_{\bot }^{-}\left( 1-\exp \left( -\xi \right) \right) \mathbf{=}
\label{Balance of forces in the Hartmann layer} \\
\mathbf{-}\dfrac{1}{\rho }\mathbf{\mathbf{\nabla }_{\bot }}p-\dfrac{\sigma }{%
\rho }\mathbf{\nabla }_{\bot }\phi \times \mathbf{B}-\dfrac{\sigma }{\rho }%
\mathbf{B}^{2}\mathbf{u}_{h}\left( \xi \right) \mathbf{+}\nu \partial
_{zz}^{2}\mathbf{u}_{h}\left( \xi \right) .
\end{multline}
Since the pressure and the electric potential do not vary at the scale of
the Hartmann layer, after a straightforward integration, we get : 
\begin{multline}
\mathbf{u}_{h}\left( \xi \right) =\mathbf{u}_{\bot }^{-}\left( 1-\exp \left(
-\xi \right) \right)   \label{Hartmann layer inertial profile} \\
+\dfrac{t_{H}}{Ha}\left[ \left( \dfrac{1}{3}e^{-2\xi }-\dfrac{1}{3}e^{-\xi
}+\xi e^{-\xi }\right) \mathbf{u}_{\bot }^{-}.\nabla _{\bot }\mathbf{u}%
_{\bot }^{-}+\dfrac{1}{2}\xi e^{-\xi }\partial _{t}\mathbf{u}_{\bot }^{-}%
\right] .
\end{multline}
It is noticeable that this profile is not divergent-free so that it implies
some flow rate between the Hartmann layer and the core flow. The vertical
velocity at the interface of the two regions $w^{-}$\ is derived from the
mass conservation : 
\begin{equation}
w_{h}^{-}=-\dfrac{5}{6}\dfrac{at_{H}}{Ha^{2}}\mathbf{\nabla }_{\bot }.\left( 
\mathbf{u}_{\bot }^{-}\mathbf{.\nabla }_{\bot }\mathbf{u}_{\bot }^{-}\right) 
\label{vertical flow rate}
\end{equation}
It should be noticed too that this behavior is not dependent upon the
electric conditions at the Hartmann walls.

\section{Application to the case of isolated vortices.}

Using the models of the previous paragraph for the velocity profile in both
the Hartmann layer and the core flow, it is possible to express $\mathbf{u}%
_{\bot }^{-}$ in function of $\mathbf{\bar{u}}$ and to assess $\mathbf{\tau }%
_{W}$ and $\overline{\mathbf{u}^{\prime }\mathbf{.\nabla }_{\bot }\mathbf{u}%
^{\prime }}$ in (\ref{2D integrated eq - general form.}) to obtain a 2D
equation accounting for 3D\ inertial phenomena. In order to get a simple
model, we limit ourselves to the inertial effects in the Hartmann layer and
we neglect the core ''barrel effect'' corrections are taken on account. It
can be checked that this becomes exact in axisymmetric configurations. Under
this assumptions, (\ref{2D integrated eq - general form.}) yields :

\begin{equation}
\dfrac{d\mathbf{v}}{dt^{\prime }}+\mathbf{\nabla }_{\bot }\bar{p}^{\prime
}=\nu \mathbf{\Delta }_{\bot }\mathbf{v-}\dfrac{1}{t_{H}}\left( \mathbf{v}%
_{0}-\alpha \mathbf{v}\right) \mathbf{+}\dfrac{t_{H}}{Ha^{2}}\left( \dfrac{7%
}{36}\mathcal{D}_{\mathbf{v}}+\dfrac{1}{8}\partial _{t}\right) \mathbf{v}.%
\mathbf{\nabla }_{\bot }\mathbf{v}  \label{2D axi equation motion}
\end{equation}
where $\alpha =1+1/Ha,$ $\mathbf{v}$ (resp. $\mathbf{v}_{0}$, resp. $\bar{p}%
^{\prime }$)$.\mathbf{=}\left( 1+7/\left( 6Ha\right) +11/6Ha^{2}\right) 
\mathbf{\bar{u}}$ (resp. $\mathbf{u}_{0}$, resp. $\bar{p}$), $\partial
t^{\prime }=\left( 1+1/Ha+n^{2}/Ha^{2}\right) \partial t$ and $\mathcal{D}_{%
\mathbf{v}}$ stands for the operator $\mathcal{D}_{\mathbf{v}}:\mathbf{%
F\longmapsto }\mathcal{D}_{\mathbf{v}}\mathbf{F=}\left( \mathbf{v.\nabla }%
_{\bot }\right) \mathbf{F+}\left( \mathbf{F.\nabla }_{\bot }\right) \mathbf{%
v.}$

This model may be checked by comparison of its results for isolated vortices
with the measurements performed by Sommeria (\cite{Sommeria88}). His
experimental device consists in a cylindrical tank (radius $60$ mm) filled
with a layer of mercury (depth $19.2$ mm) at the center of which an electric
current is injected \textit{via} a point electrode located at the bottom.
The upper surface is free so that the velocities are derived from streak
photos of particles conveyed by the fluid.

In this configuration, if $I$ is the injected electric current, the forcing
expresses as (\cite{Sommeria88}) : 
\begin{equation}
u_{0\theta }=\dfrac{\Gamma }{r}\text{ \ with \ }\Gamma =\dfrac{I}{2\pi \sqrt{%
\rho \sigma \nu }}.  \label{pointnelectrode circulation}
\end{equation}
so that the equation can be rewritten in steady axisymmetric configuration
and using  non-dimensional variables $r_{\dim }=r\sqrt{\nu t_{H}/\alpha }$
and $u_{\theta \dim }=u_{\theta }\Gamma /a\sqrt{\alpha Ha}$ :
\begin{equation}
C_{t}\dfrac{1}{r^{2}}\partial _{r}\left( ru_{\theta }^{3}\right) =\dfrac{1}{%
r^{2}}\partial _{r}\left[ r^{3}\partial _{r}\left( \dfrac{u_{\theta }}{r}%
\right) \right] -u_{\theta }+\dfrac{1}{r},  \label{non lin tourb adim eq}
\end{equation}
with the corresponding boundary conditions :

\begin{equation}
u_{\theta }\left( R\right) =0\text{and}\underset{r\rightarrow +\infty }{\lim 
}u_{\theta }=0.  \label{non lin tourb boundary conditions}
\end{equation}
In (\ref{non lin tourb adim eq}) $C_{t}=\dfrac{7}{36}\left( 1+\dfrac{1}{Ha}%
\right) ^{3/2}\dfrac{1}{N_{c}^{2}}$ where $N_{c}=\dfrac{\sigma B^{2}}{\rho }%
\dfrac{a^{2}}{\Gamma Ha}$ , the interaction parameter scaled on the solid
core of the vortex (\cite{Sommeria88}) appears as the relevant parameter for
this problem. We have performed numerical simulations of (\ref{non lin tourb
adim eq}) for two different values of the injected current ($I=50mA$ and $%
I=200mA$ respectively corresponding to $C_{t}=8.85$ and $C_{t}=141.65$ ).
The results are reported on figure (\ref{Vortex radial profile of angular
momentum}). A satisfactory agreement is found between theory and experiment.
It turns out that the core of the vortices actually broadens for increasing
values of the injected current (\textit{i.e.} when inertia becomes
important). This phenomenon can be interpreted as an Ekman recirculation :
the flow rate between the Hartmann layer and the core (\ref{vertical flow
rate}) is related to the spatial variation of the centrifugal acceleration
of the fluid, which is strong and positive at the center of the vortex \
(strong flow towards the core) and negative and weak at long distance of the
center. This toroidal flow is an MHD equivalent of the Ekman pumping.

\begin{figure}
\centering
\includegraphics[width=0.8\textwidth]{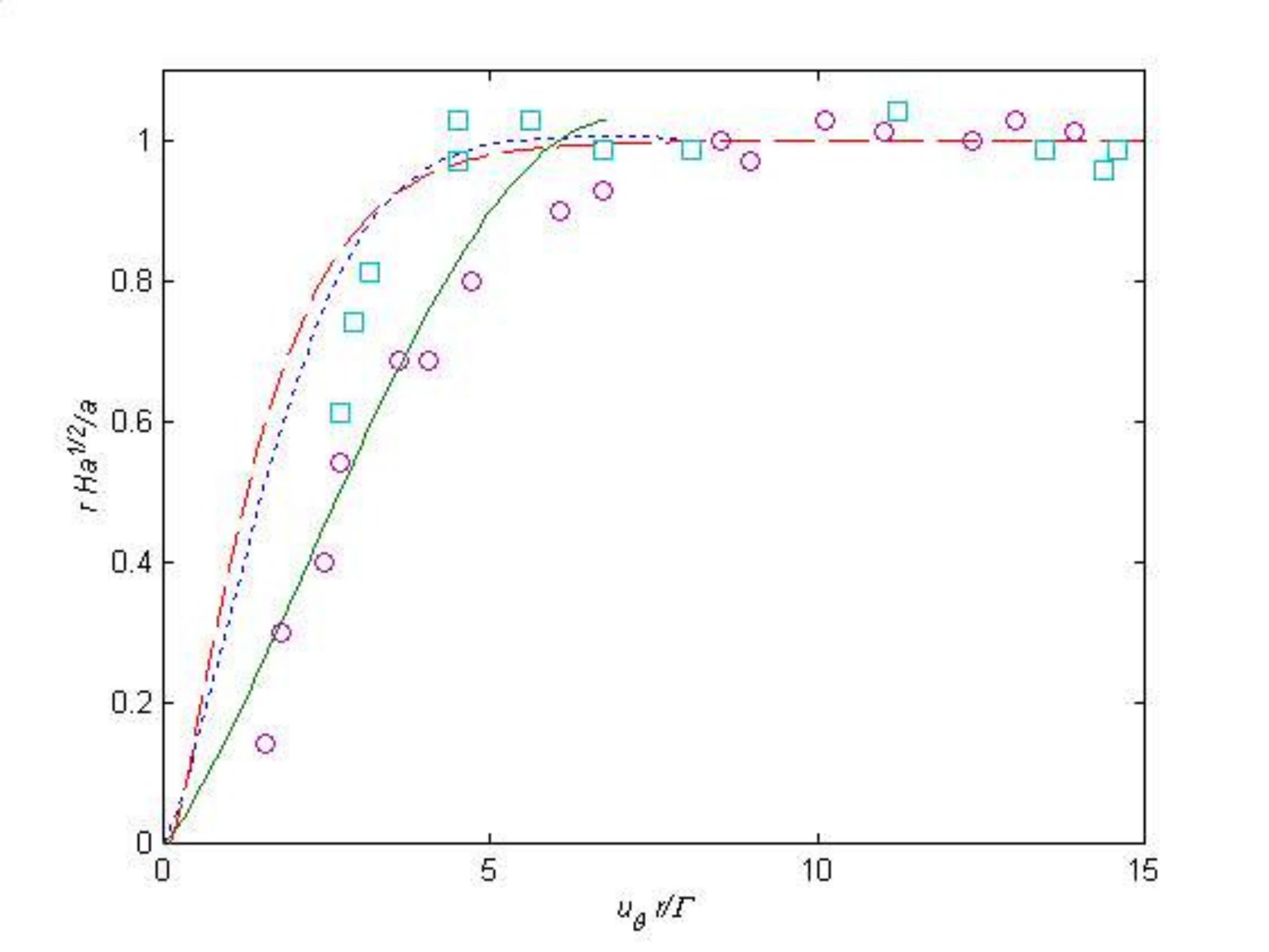}
\caption{Vortex radial profile of angular momentum for $B = 0.5$ T: boxes: experimental measurements for injected current $I = 0.05$ A, circles: experimental measurements for $I = 0.2$ A, full line: analytical profile 
without non linear effects, semi dotted line: numerical profile for $I = 0.05$ A, dotted line: numerical 
profile for $I = 0.2$ A. Note that numerical precision problems don’t allow to get to profiles for any values of r.
\emph{Profil radial de moment cin\'{e}tique de tourbillons isol\'{e}s pour 
$B=0.5T$ : (carr\'{e}s) : Mesures exp\'{e}rimentales pour un courant de $I=0.05A,$ ronds : mesures
ep\'{e}rimentales pour $I=0.2A,$ ligne pleine : courbe analytique sans effet non lin\'{e}aire, ligne interrompue : courbe num\'{e}rique \`{a} $I=0.05A$, pointill\'{e}s : courbe num\'{e}rique \`{a} $I=200A$.}
\label{Vortex radial profile of angular momentum}
}
\end{figure}



\end{document}